\documentclass[prb,showpacs,twocolumn,superbib,floats,,superscriptaddress]{revtex4}
\usepackage{graphicx}
\usepackage{epsfig}
\usepackage{amsfonts}
\usepackage{amsmath}
\usepackage{natbib}
\usepackage{multirow}
\usepackage{bm}
\usepackage{epstopdf}
\usepackage{color}
\usepackage{ulem}

\begin{document}

\title{Semiempirical pseudopotential approach for nitride-based 
nanostructures and {\it ab initio} based passivation of free surfaces}

\author{A. Molina-S\'{a}nchez}
\affiliation{Max-Planck Institut f\"{u}r Festk\"{o}rperforschung, Stuttgart, Germany}
\affiliation{Instituto de Ciencia de los Materiales de la Universidad de Valencia, E-46071 Valencia, Spain}
\affiliation{Physics and Material Sciences Research Unit, Universit\'e du Luxembourg, Luxembourg}
\author{A. Garc\'{i}a-Crist\'{o}bal}
\affiliation{Instituto de Ciencia de los Materiales de la Universidad de Valencia, E-46071 Valencia, Spain}
\author{G. Bester}
\affiliation{Max-Planck Institut f\"{u}r Festk\"{o}rperforschung, Stuttgart, Germany}

\date{\today}

\begin{abstract}
We present a semiempirical pseudopotential method based on screened atomic pseudopotentials and derived 
from \textit{ab initio} calculations. This approach is motivated by the demand for pseudopotentials able to 
address nanostructures, where \textit{ab initio} methods are both too costly and insufficiently accurate at
the level of the local-density approximation,
while mesoscopic effective-mass approaches are inapplicable due to the small size 
of the structures along, at least, one dimension.
In this work we improve the traditional pseudopotential method by a two-step process: First, we 
invert a set of self-consistently determined screened {\it ab initio} potentials in wurtzite GaN for 
a range of unit cell volumes,
thus determining spherically-symmetric and structurally averaged atomic potentials.
Second, we adjust the potentials to reproduce observed excitation energies. We find
that the adjustment represents a reasonably small perturbation over the potential, so that
the ensuing potential still reproduces the original wave
functions, while the excitation energies are significantly improved.
We furthermore deal with the passivation of the dangling bonds of free surfaces which is relevant for the study of nanowires and colloidal nanoparticles. We present a
methodology to derive passivant pseudopotentials from {\it ab initio} calculations. We apply 
our pseudopotential approach to the exploration of the confinement effects on the 
electronic structure of GaN nanowires.
\end{abstract}


\maketitle

\section{Introduction}
\label{intro}

Much of the computational effort in solving the Kohn-Sham equations of density 
functional theory (DFT)\cite{hohenberg64a,kohn65a} 
is spent in iteratively updating the screened effective potential. This 
procedure requires the calculation of the electronic density, obtained over a sum of all 
occupied Kohn-Sham eigenstates. The required number of states (bands) therefore 
scales with the number of atoms. For the calculation of optical or transport 
properties, only the bands close to the band gap are required, which represents,
for a large number of atoms, only a small fraction of the total number of bands required for the calculation 
of the total energy in a self-consistent formalism. An approach that bypasses 
the calculation of the entire spectrum and focuses on states close to the band 
gap is the semi-empirical pseudopotential approach.\cite{wang95,bester09} In this 
approach the pseudopotential is derived from a DFT calculation in the local density 
approximation (LDA) and subsequently empirically corrected to reproduce 
experimental {\it target} properties, such as the optical band gap. These 
semi-emipirical pseudpotentials (SEPs) were derived for CdSe and InP and 
their accuracy demonstrated.\cite{wang95,fu97b} In this work, we follow a 
similar procedure for the derivation of nitride SEPs. Former treatment of 
nitrides were at the level of empirical pseudopotentials (as opposed to SEPs) 
which assume an empirical analytic functional dependence (a sum of Gaussians) 
instead of being obtained from DFT-LDA and include only a local component, while 
the SEPs are angular momentum dependent non-local (or semi-local)  
potentials.\cite{bellaiche96,grundmann} These empirical pseudopotentials have a 
parametric strain dependence that is important and more intricate in the case of nitrides than 
for the case of III-V semiconductors.\cite{williamson00}  Our derivation of SEPs for 
nitrides leads to pseudopotentials that are slightly more expensive 
computationally (they have an angular momentum dependent semi-local 
component and a slightly larger energy cut-off) but more accurate and 
transferable. There is no need for an explicit empirical parametric 
strain dependence. The advantages of SEPs compared to the local empirical pseudopotentials 
have been discussed earlier, generally to zinc-blende 
semiconductors.\cite{wang95,fu97b,bester09} In this work we have extended the SEP formulation
to wurtzite semiconductors, and more specifically to GaN.	

The second development that was necessary to properly address nitride nanostructures 
was the problem of the passivation. In earlier work \cite{wang95,graf07} the dangling
bonds of nanostructures were passivated by empirical Gaussian potentials. The three 
parameters describing the passivant, namely the width and height of the Gaussian and 
its distance to the passivating atom, were varied until the surface states, that fill 
the gaps of most unpassivated semiconductors, disappears. This fitting procedure was 
originally done manually and subsequently automatized using a minimization 
procedure.\cite{graf07} Even in the case of the exhaustive numerical search of the 
configuration space, some materials could not be properly 
passivated. This fact is probably due to the lack of flexibility in the Gaussian 
function used. In this paper we describe our approach to obtain passivant 
potentials. It is based on the DFT-LDA calculation of a passivated and an 
unpassivated (bare) slab. From the difference of both calculations we extract 
an effective passivant potential in real space. We can accurately fit the numerical 
results to a Yukawa-type potential. 

Once the bulk SEPs for a given material are found, together with its corresponding
passivants pseudopotentials, the study of the electronic structure and related
properties of nanostructures can be performed. We have chosen as an example to apply the 
SEP method to GaN nanowires,\cite{Goldberger} a promising
nanostructure in the field of solar cells and optoelectronic devices due to the wide band gap of 
GaN (3.5 eV), that together with indium nitride, a narrow band gap semiconductor (0.67 eV) can cover
the full visible solar spectrum by a proper alloying.\cite{wu} More specifically, we explore the 
effects of confinement in the valence and conduction band states close to the nanowire band gap and
we discuss the influence of confinement on the optical properties.

\section{Construction of semiempirical pseudopotentials from bulk LDA calculations}
\label{theory}

First, we solve the Kohn-Sham equations for the bulk wurtzite structure 
within the LDA using standard {\it ab initio} norm-conserving nonlocal 
pseudopotentials.\cite{gonze09} 

\begin{equation}
\left[ -\frac{1}{2}\nabla^2+V_\mathrm{KS}\right] \psi_i(\bm{r}) =\varepsilon_i\psi_i(\bm{r}).
\label{kseq}
\end{equation}

The Kohn-Sham potential $V_\mathrm{KS}$ includes the local and nonlocal ionic pseudopotentials and the
electron-electron interaction:

\begin{equation}
V_\mathrm{KS} = V_\mathrm{local}(\bm{r}) + V_\mathrm{nonlocal}(\bm{r}) + V_\mathrm{HXC}(\bm{r}) 
\end{equation}

The ensuing local part of the self-consistent screened 
effective potential is defined as:
\begin{equation}
V_\mathrm{LDA}(\bm{r})=V_\mathrm{local}(\bm{r}) + V_\mathrm{HXC}(\bm{r}),
\end{equation}
where $V_\mathrm{local}(\bm{r})$ is the local part of the non-screened ionic 
pseudopotential and the term $V_\mathrm{HXC}(\bm{r})$ includes the Hartree 
and exchange-correlation contributions. A more detailed discussion concerning the 
nonlocal part of the pseudopotential can be found elsewhere.\cite{Parr} The 
potential $V_\mathrm{LDA}(\bm{r})$ is a periodic function and 
can be expanded in a Fourier series:
\begin{equation}\label{fourierLDA}
V_\mathrm{LDA}(\bm{G})=  \int_{\Omega}d^3\bm{r}\,V_\mathrm{LDA}(\bm{r})\,e^{-i\bm{G}\cdot\bm{r}}
\quad ,
\end{equation}
where $\{\bm{G}\}$ is the set of the reciprocal lattice vectors of
the corresponding crystal structure with unit cell volume $\Omega$.
 
We now express the periodic potential $V_\mathrm{LDA}(\bm{r})$
as a sum over atom-centered potentials $v_{\alpha}(\bm{r})$:
\begin{equation}
V_\mathrm{LDA}(\bm{r}) =
\sum_{\alpha}\sum_{\bm{R}}
v_{\alpha}(\bm{r}-(\bm{R}+\bm{\tau}_{\alpha}))\quad ,
\label{atom_centered}
\end{equation}
where we use capital letters for crystal potentials and lower case letters for 
the corresponding atomic
potentials. The vectors $\{\bm{R}\}$ are the basis vectors of the crystal unit 
cell and
$\bm{\tau}_{\alpha}$ are the atomic positions of the atoms $\alpha$ in the unit 
cell corresponding to $\{\bm{R}\}$.

The goal now is to obtain an expression for
$v_{\alpha}$ that we could use in other environments, as give for 
instance in nanostructures. These
\textit{screened} atomic pseudopotentials $v_{\alpha}$ contain,
in addition to the ionic part, in an average way, the contribution
related to the electron-electron interaction, expressed by
$V_{\rm{HXC}}$. However, the procedure of obtaining these atomic
potentials from the total local potential must be clearly specified
and checked for consistency. In the following, we explain how the
construction of the screened atomic pseudopotential is carried out.

First, by combining Eqs.~\eqref{fourierLDA} and \eqref{atom_centered}, we obtain the following
connection between the unknown $v_{\alpha}(\bm{G})$ and the known self-consistent
Fourier coefficients $V_\mathrm{LDA}(\bm{G})$, for every value of $\bm{G}$:
\begin{equation}
\sum_{\alpha}\,e^{-i\bm{G}\cdot\bm{\tau}_\alpha}\,
v_{\alpha}(\bm G)=V_\mathrm{LDA}(\bm G),
\end{equation}
with
\begin{equation}
v_{\alpha}(\bm{G})=\int_{\Omega_0}v_{\alpha}(\bm{r})e^{-i\bm{G}\cdot\bm{r}}d^3\bm{r},
\end{equation}
where the integral is over the atomic volume $\Omega_0$, as a consequence of 
Eq.~\eqref{atom_centered} and after
Ref.~\onlinecite{bester09}. The lattice parameters are taken as
$a=3.189$ \AA, and $c=\sqrt{8/3}a$. The wurtzite unit cell has four 
atoms, with the primitive vectors
\begin{equation}
\begin{aligned}
\bm{a}_1 & = \left(\frac{a}{2},\frac{\sqrt{3}a}{2},0\right),   \\
\bm{a}_2 & = \left(-\frac{a}{2},\frac{\sqrt{3}a}{2},0\right),  \\
\bm{a}_3 & = \left(0,0,c\right) \quad ,                               
\end{aligned}
\label{primitive_vectors}
\end{equation} 
and the atom positions of the two anions $a_{1,2}$ and two cations $c_{1,2}$
\begin{equation}
\begin{aligned}
\bm{\tau}_{a_1}&=-\frac{1}{6}\bm{a}_1+\frac{1}{6}\bm{a}_2-\frac{7}{16}\bm{a}_3
\longrightarrow   \bm{\tau}_{a}, \\[2pt]
\bm{\tau}_{a_2}&=+\frac{1}{6}\bm{a}_1-\frac{1}{6}\bm{a}_2+\frac{1}{16}\bm{a}_3\longrightarrow
-\bm{\tau}_{c},\\[2pt]
\bm{\tau}_{c_1}&=-\frac{1}{6}\bm{a}_1+\frac{1}{6}\bm{a}_2-\frac{1}{16}\bm{a}_3\longrightarrow
\bm{\tau}_{c},  \\[2pt]
\bm{\tau}_{c_2}&=+\frac{1}{6}\bm{a}_1-\frac{1}{6}\bm{a}_2+\frac{7}{16}\bm{a}_3\longrightarrow  -\bm{\tau}_{a},\\
\end{aligned}
\end{equation}
we then obtain for  $V_\mathrm{LDA}(\bm{G})$:
\begin{equation}
\begin{aligned}
V_\mathrm{LDA}(\bm{G})=\frac{1}{\Omega_c}[e^{-i\bm{G}\cdot\bm{\tau}_a}v^{(a)}(\bm{G}) + e^{-i\bm{G}\cdot\bm{\tau}_c}v^{(c)}(\bm{G})+ \\
e^{ i\bm{G}\cdot\bm{\tau}_c}v^{(a)}(\bm{G}) + e^{ i\bm{G}\cdot\bm{\tau}_a}v^{(c)}(\bm{G})],
\end{aligned}
\end{equation}
with the symmetric ($v_{+}$) and antisymmetric ($v_{-}$) pseudopotentials
\begin{eqnarray}
v_+(\bm{G})&=&v_a(\bm{G})+v_c(\bm{G})\quad, \\ \nonumber
v_-(\bm{G})&=&v_a(\bm{G})-v_c(\bm{G}),
\label{def_vpvm}
\end{eqnarray}
we obtain
\begin{equation}
\begin{aligned}
V_\mathrm{LDA}(\bm{G}) & = & \frac{1}{\Omega_c}v_+(\bm{G})\left[\cos\left(\bm{\tau}_a\cdot\bm{G}\right)+\cos\left(\bm{\tau}_c\cdot\bm{G}\right)\right]\\
\\
& + & i\frac{1}{\Omega_c}v_-(\bm{G})\left[\sin\left(\bm{\tau}_c\cdot\bm{G}\right)-\sin\left(\bm{\tau}_a\cdot\bm{G}\right)\right].\\
\end{aligned}
\end{equation}

Since $v_{a/c}$ have inversion symmetry, $v_{+/-}$ are real, and
they can be written in terms of the real and imaginary parts of
$V_\mathrm{LDA}(\bm{G})$

\begin{equation}
\begin{aligned}
v_+(\bm{G}) & = & \frac{\Omega_c}{\cos\left(\bm{\tau}_a\cdot\bm{G}\right)+\cos\left(\bm{\tau}_c\cdot\bm{G}\right)}\Re\{V_{\rm LDA}(\bm{G})\},\\
\\
v_-(\bm{G}) & = &
\frac{\Omega_c}{\sin\left(\bm{\tau}_c\cdot\bm{G}\right)-\sin\left(\bm{\tau}_a\cdot\bm{G}\right)}\Im\{V_{\rm
LDA}(\bm{G})\}. \label{def_vpvm2}
\end{aligned}
\end{equation}

The potential $V_{\rm LDA}(\bm{G})$ is calculated from the Fourier transform of the LDA crystal potential, $V_{\rm LDA}(\bm{r})$. Once $v_{+/-}$ 
are obtained, the reciprocal-space atomic
pseudopotentials $v_{a/c}$ are obtained from Eq.~(11).
Through this procedure we can obtain $v_{a/c}(\bm{G})$ only at discrete values
$\{\bm{G}\}$. To palliate this problem the LDA potential of the bulk system is
solved for a set of unit cell volumes, at different lattice constants, around the 
optimized value, as shown below.

So far the procedure to obtain $v_{(a/c)}(\bm{G})$ is exact. However, for an 
efficient implementation it is convenient to assume that the screened pseudopotentials 
are well represented by their spherically averaged counterparts $v^{(a/c)}(G)$. 
At this stage we have a spherically symmetric screened atomic LDA potential
$v_{a/c}(G)$. 

\begin{figure}
\includegraphics[width=7.0 cm]{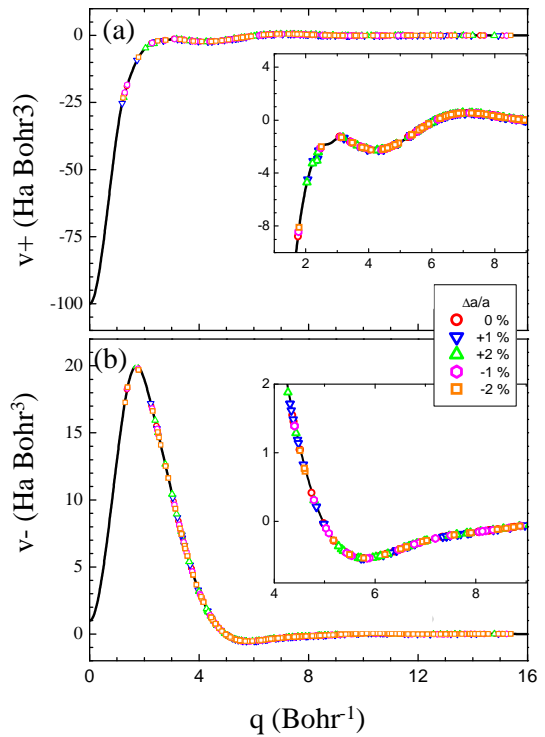}
\caption{The LDA-derived contributions $v_+$ and $v_-$ for GaN as 
defined in the main text. We have obtained five sets of data points by varying 
the lattice constant $a$ as indicated in the legends. The $c/a$ 
ratio is kept constant at the ideal value. The lines 
represent a fit to the date points using cubic splines.}
\label{vplusminus}
\end{figure}

Figures~\ref{vplusminus}(a) and \ref{vplusminus}(b) show the
LDA results (symbols) for $v_{+}$ and $v_{-}$.
In order to have a dense grid of $\bm{G}$ points the calculations are performed for five different lattice constants $a$.  
The data points from different unit cell volumes fall nearly over the same line, with some slight
deviations for the data points around $G = 2$~Bohr$^{-1}$ in $v_{+}$.
This shows that the spherical approximation is justified. 
Concerning the pseudopotentials at $\bm{G}=0$, $v_{+/-}(\bm{G}=0)$,
their values are arbitrary, and the variation of $v_{+}(\bm{G}=0)$
produces a overall shift of the band structure. The value of
$v_{+}(\bm{G}=0)$ is set to the work function, that can be obtained from
experimental data\cite{Grabowski2001} or obtained from 
\textit{ab initio} 
calculations. In nanostructures, as superlattices or quantum wells, $v_{+}(\bm{G}=0)$ determines 
the band alignment and its value has to be
set carefully.\cite{zhang1993,mader}  
In regions of small $G$ ($\lesssim2.5$ Bohr$^{-1}$), the changes in the
pseudopotentials $v_+$ and $v_-$ do not substantially modify the
band structure around the $\Gamma$ point. In all the cases, we have fitted the data points with cubic splines.
In regions of large $G$ ($>12$ Bohr$^{-1}$), small oscillations persist, requiring a careful fitting. 
\begin{figure}
\includegraphics[width=7.0 cm]{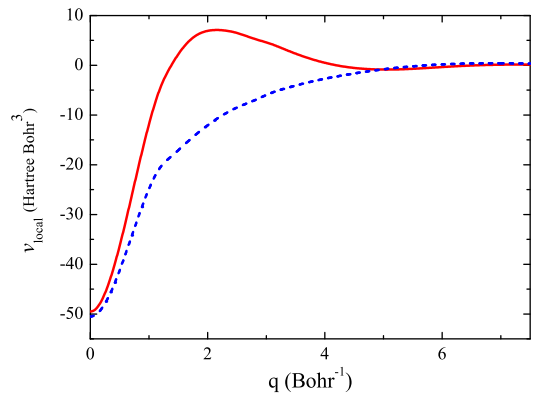}
\caption{Screened atomic semiempirical pseudopotentials for 
gallium (solid line) and nitrogen 
(dashed line). 
}
\label{bulkpot}
\end{figure}
In Fig.~\ref{bulkpot} we plot the corresponding gallium and
nitrogen semiempirical pseudopotentials. Concerning the general behavior of 
the pseudopotentials, the oscillations persist until $G\approx10$~Bohr$^{-1}$, which determines 
the energy cut-off in
the subsequent calculations (with a energy cutoff of 30 Ry). 

The last step is the solution of the Kohn-Sham equation~\eqref{kseq}, where the Kohn-Sham
potential $V_\mathrm{KS}$ is replaced by the
semiempirical pseudopotentials calculated above, and the nonlocal term $V_{\rm nonlocal}$.
This equation needs to be solved only once, since there is no need to achieve self-consistency. 

\begin{figure}
\includegraphics[width=7.0 cm]{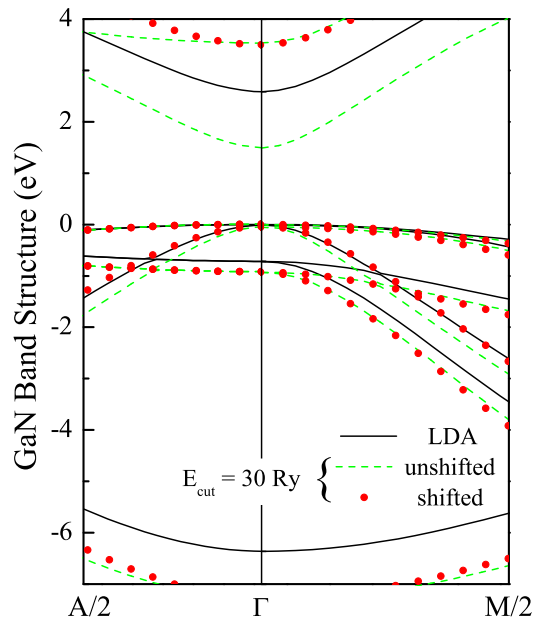}
\caption{Band structure of GaN calculated with DFT in the 
LDA (solid lines), with the SEP method without correcting the band gap (dashed line), and
with the SEP method correcting the band gap (dots). $A/2$ and $M/2$ indicates the path 
in the Brillouin zone. The LDA band-gap is 2.7 eV, 1.6 for unshifted SEP and 3.5 for shifted SEP.
}
\label{sepbulk}
\end{figure}

We proceed by comparing the band structures calculated using the SEP method and the LDA. As
the spin-orbit interaction only produces minor splittings in GaN, we
have neglected this interaction for the overall band structure
comparison, but it will be introduced later on, once the validity of
the SEP method has been demonstrated. The effects on the band structures
caused by a reduction of the energy cut-off will also be analyzed. In
Fig.~\ref{sepbulk} we plot the following GaN band structures: the LDA results for 
$E_{cut}=90$ Ry (black solid lines), the SEP results with 
$E_{cut}=30$ Ry, (green dashed lines) and the band gap-corrected SEP results 
(red dots).

A comparisons of LDA and SEP calculations with the same energy cut-off shows a 
nearly perfect agreement for all the bands and throughout the
Brillouin zone and is not shown. The reduction of the enegy cut-off from 90 Ry to 30 Ry 
lead to a rigid displacement of the conduction bands towards lower energy, reducing the band
gap approximately by 1 eV. The valence band states most
affected by the reduction of $E_{cut}$ are those far from the top of 
the valence band, at around $-6$ eV. The topmost valence band states remain
almost unaltered, and belong to the 
$\Gamma_{5}$ representation, in agreement with the \textit{ab initio} band
structure. We conclude that the LDA and the SEP band structures are in good agreement
for the bands close to the band gap, i.e., the most
important part of the band structure in, e.g., photoluminescence experiments.

Another important issue in \textit{ab initio} calculations of semiconductors
is the underestimation of the band gap undermining the LDA.\cite{bandgap} Such
underestimation is enhanced, in this special case, if the
energy cut-off is reduced. In the SEP-method one can correct this problem
easily by multiplying the local potentials by a
Gaussian function or  renormalizing the nonlocal pseudopotentials.\cite{wang95} In this work we
have chosen the second possibility where we act on the non-local angular momentum $s$-channel. Figure~\ref{sepbulk} shows the 
band structure of the band gap corrected SEPs. 
The band gap opening does not influence significantly
the valence band, although a slight modification is
appreciated in the vicinity of the $\Gamma$ point, which produces a change in the curvature, as expected from
the decreased coupling with the valence bands. The conduction band effective mass for the corrected 
SEPs is now $0.21 m_0$, in good agreement with the reported 
experimental value of $0.20 m_0$ (being $m_0$ the free-electron mass).\cite{monemar}

For energy cut-offs lower than 30 Ry, the changes in the band  structure exceed the admissible margin. 
The wave functions of the top of the valence band and of the bottom of 
the conduction band have the same symmetry and shape
than the LDA wave functions.

\section{Surface states and passivation}
\label{surface}

As noted above, the atoms at the free surfaces of a nanostructure
have dangling bonds. The existence of
these dangling bonds generates electronic states localized at the
surface, with energies usually within the band gap of the
semiconductor.\cite{wang96a,franceschetti99}
These states can have energies close
to the conduction and valence band edges, and can induce an
artificial modification of the conduction and valence band states,
and they have to be eliminated.\cite{puzder03,puzder04a,puzder04b}
In the literature two methods are used to overcome
this difficulty. The first possibility is to achieve quantum
confinement by embedding the structure into a virtual material with a 
high band gap \cite{califano03,baskoutas11}. 
The main disadvantage of using this virtual material is the associated increase
in the number of atoms.
Another way consists of attaching an atom,
usually hydrogen, to the dangling bonds (the so-called passivation),
which displaces the energies of the surfaces states far from the
band gap.\cite{xu93,huang05,reboredo01} However, the determination of the hydrogen
pseudopotentials is not a trivial task, and it can depends on the
material and of the surface orientation. The simplest way is to
model the passivant pseudopotential by an analytical function, such
as a Gaussian. The method to
optimize the form of the Gaussian function can be a simple trial and
error procedure, or a more sophisticated approach, such as a
genetic algorithm, as developed in Ref.~\onlinecite{graf07}. However, the
empirical parametrized Gaussian pseudopotentials involve iterative
calculations of the electronic structure, which complicates the
computational task, and does not always guarantee to achieve an
adequate passivation.

In this work, we propose to derive the passivant pseudopotentials
from LDA calculations of a small free-standing structure.
In the following we describe the necessary steps and illustrate them by an 
example for the $(1\bar{1}00)$ surface of GaN.

\textit{Step 1: Ab initio calculation of the self-consistent potential.} 
The self-consistent potential is calculated for two free-standing
$(1\bar{1}00)$ GaN layers. A bare layer (unsaturated dangling bonds) 
and a passivated layer (a hydrogen attached to each dangling bond). Specifically, the GaN layer
is approximately 15 angstrom thick (24 atoms), and the vacuum layer is about 10 angstrom on both sides
to prevent coupling between layers.
The input hydrogen norm-conserving potential is taken to be the same for
the hydrogen atoms bound to the Gallium and the Nitrogen atom. In the case of materials with 
metallic surfaces (e.g. CdSe), a satisfactory passivation may require hydrogen potentials
with fractional charges. For the atomic positions, we fix the Nitrogen and Gallium atoms 
to their bulk ideal positions 
and relax only the distance between the hydrogen atoms (usually two atoms on each side of the 
slab) and the corresponding passivated 
atom.\cite{distances}

\begin{figure*}
\includegraphics[width=15 cm]{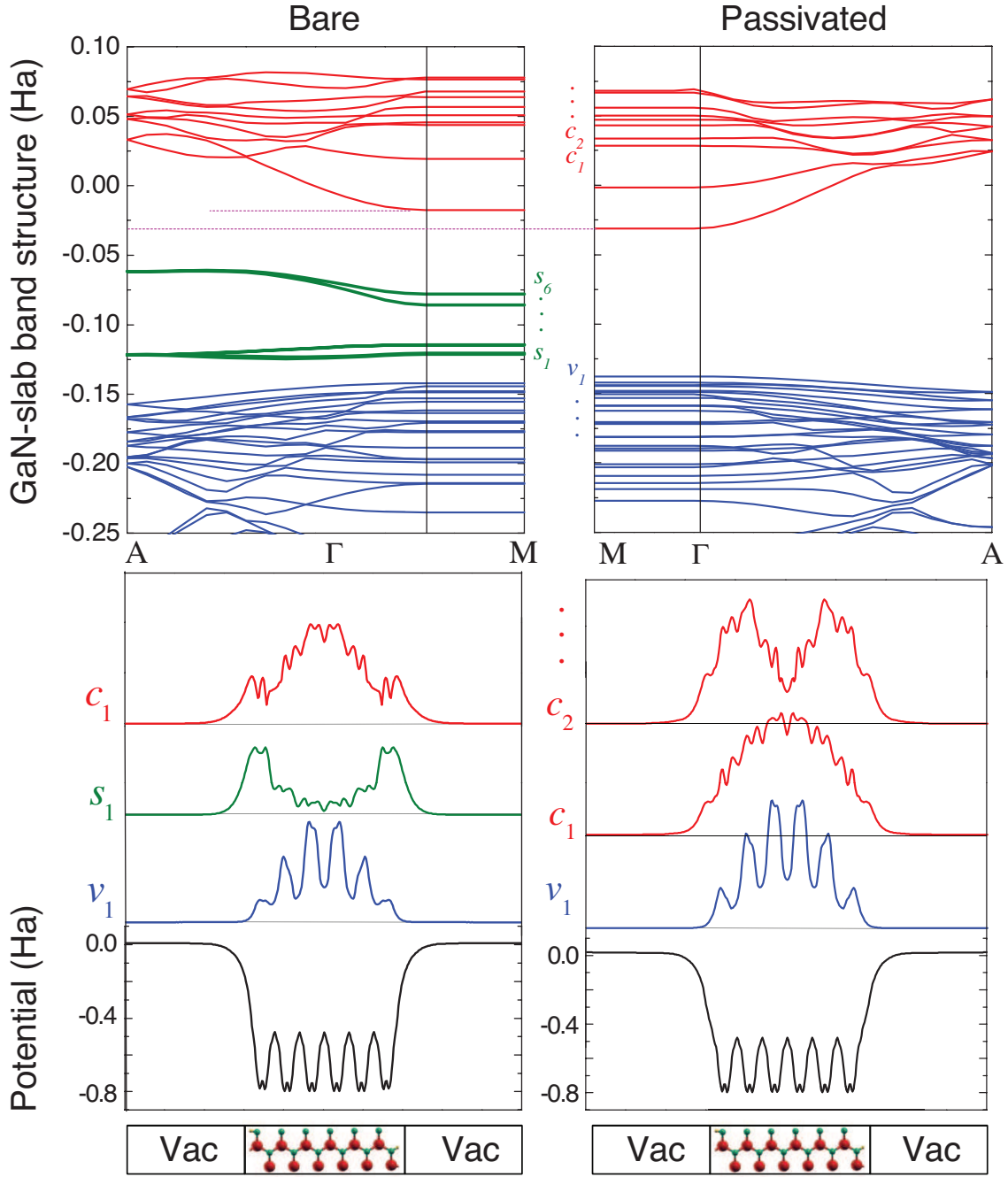}
\caption{Top panels: Band structure of a bare and passivate $[1\bar{1}00]$ GaN 
slab calculated with DFT in the LDA. Middle panels: Square of the wave functions 
of selected conduction ($c_1$, $c_2$), valence ($v_1$) and surface ($s_1$) states, along 
with the screened local effective DFT-LDA potential averaged along $(1\bar{1}00)$ 
planes.  Lower panels: schematic representation of the atomic positions. .
}
\label{ABO_layer}
\end{figure*}

The band structures of the bare slab (without geometry optimization) 
and passivated $[1\bar{1}00]$ slab (with
optimization of the distance of hydrogen atoms with respect to GaN) 
are shown in Figs.~\ref{ABO_layer}(a) and 
(b). Note that the bands are rather flat along the $\Gamma-M$ direction, which confirms
the decoupling between adjacent layers. The energy dispersion along 
the $\Gamma-A$ line of the Brillouin zone is shown, exhibiting the same behavior as 
in the case of a quantum well.\cite{molina09} We observe that conduction and valence bands
are very similar in both cases. However, the bare layer has a band gap larger by 0.37 eV. The 
main difference is obviously 
the presence of surface states, located within the band gap. These states are grouped in two energy 
ranges. The states $s_1-s_4$ (index assigned in increasing order of energy), are closer to the valence 
band edge and have a flatter dispersion. The other group, formed by the states $s_5$ and $s_6$, located in the middle of the 
band gap, have a more pronounced dispersion. On the contrary, the band structure of the passivated layer has no state within the band gap.

This analysis is further clarified by examining the charge density in real space. Figures~\ref{ABO_layer}(c) and (d) 
show the plane-averaged density corresponding to some relevant states of the bare and passivated $(1\bar{1}00)$-layers, together
with the plane-averaged self-consistent potential profile, along the $[1\bar{1}00]$ direction. In the case of the bare layer, the
mid-gap surface states $s_5$ and $s_6$ present different profiles having a strong hybridization with the layer 
atoms, penetrating appreciably inside the layer. For the state $c_1$ the penetration into vacuum is significant for 
both, the passivate and the bare layers, which can be explained by the small electronic effective mass. In 
the case of the bare layer, the state $c_1$ develops a maximum at the surface of the layer, which is the 
signature of a hybridization with a surface state. On the other hand, the charge density of the valence 
band states seems to vanish at the surface without any sign of hybridization, in either the bare or the passivated layer. It is 
worth to mention that for larger passivated nanostructures the charge density at the border will be negligible. We can also conclude 
that the dangling bonds affect mainly the states of the conduction band.

\textit{Step 2: Determination of the screened hydrogen pseudopotential.} 
We calculate the effective pseudopotential for the atom positions of the unpassivated slab using our semi-empirical 
pseudopotentials for Ga and N, $V_{bare,SEPM}(\bm r)$. We use the same FFT grid as for the self-consistent 
potential for the passivated slab calculated in the previous step via LDA, $V_{p,LDA}$. The difference between the two sets of data
\begin{equation}
\Delta V(\bm r) = V_{p,LDA}(\bm r) - V_{bare,SEPM}(\bm r),
\end{equation}
represent the modification of the potential in response to the presence of the passivated surface. This 
includes the effects of the self-consistent charge redistribution around the surface Ga and N atoms, as well 
as the additional hydrogen potential. The potential 
$\Delta V(\bm r)$ is therefore a potential that represents the entire modification of the total potential 
trough the introduction of a passivated surface. Accordingly, we obtain a different 
passivant ``hydrogen" potential for a Hydrogen atom attached to a Ga atom and for a Hydrogen atom attached to a 
N atom. To extract a spherically averaged atomic quantity from $\Delta V(\bm r)$ we calculate the 
center of potential (akin a center of mass) $\bm r_{\rm cop} $ of a set of $N$ points enclosed within a 
sphere around the hydrogen atomic positions $\bm r_{\rm H}$:
\begin{equation}
\label{eq:rh}
\bm r_{\rm cop} = \sum_i^N  \frac{\Delta V(\bm r_i) \bm r_i}{\Delta V(\bm r_i)} \quad \mbox{for}\quad |\bm r_i - \bm r_{\rm H} | < r_{\rm cut} \quad .
\end{equation}
If the potential would be spherical and centered around $\bm r_{\rm H}$ then $\bm r_{\rm cop}$ and $\bm r_{\rm H}$ 
would coincide. In our case, the geometric potential center is slightly shifted from the atomic position and we 
solve the equation self-consistently until $\bm r_{\rm H} = \bm r_{\rm cop}$. For the value of $r_{cut}$ we use a value of 1.0 a.u..

\begin{figure}
\includegraphics[width=7.0 cm]{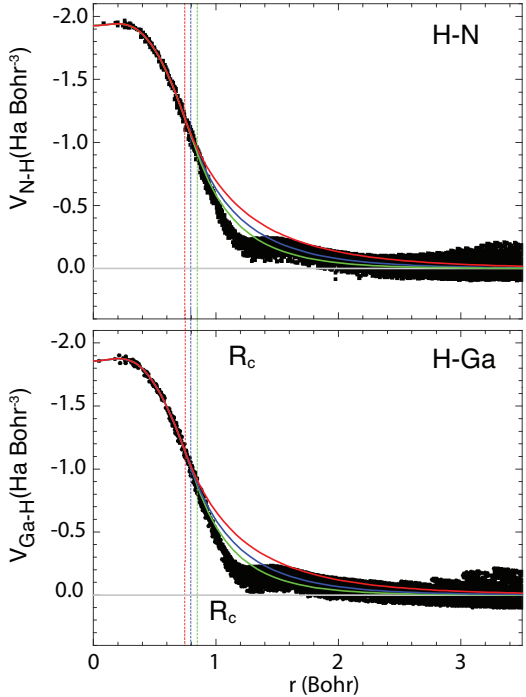}
\caption{Potential given by Eq.~(\ref{eq:passfit}) for a 15 {\AA} thick GaN film. The 
three different fits to the data points represent three different 
values of $R_C = 0.75$, $0.80$ and $0.85$ a. u. (where the cubic splines are 
connected to the Yukawa potential). We 
obtain the best results with $R_C$ = 0.80 a.u.}
\label{fig:scattered}
\end{figure}

Our passivant pseudo-hydrogen pseudopotential is given by 
\begin{equation}
\label{eq:passfit}
v_{\rm H}(r_i) = \Delta V (|\bm r_i - \bm r_{\rm H}|) \quad ,
\end{equation}
and is shown in Fig.~\ref{fig:scattered}, where we plot the discrete set of FFT grid points 
as black circles.  The position of the hydrogen atom $\bm r_{\rm H}$ has been determined from 
Eq.~(\ref{eq:rh}). The smooth curves in Fig.~\ref{fig:scattered}  represent our fits to the data 
points, where we used three cubic spline functions for $r < R_C$. In this region, the data points 
lie mainly on a curve, which shows that the potential is close to spherically symmetric. For 
larger radii, the data points show significant scattering along the enegy axis, which is 
expected, since the potential at the surface will have some non-spherical character. We obtain 
good results when the data points
are fitted by a Yukawa potential:
\begin{equation}
\label{eq:yuk}
v_{\rm H}(r) = -g^2 \frac{{\rm e}^{-m r}}{r} \quad \mbox{for}\quad r > R_C \quad .
\end{equation}

For the N (Ga) passivant potential we 
determine $m$ =  1.864344 (2.012838) and $g^2$ =  3.7635 
(4.060362) with $R_C$ = 0.80 a.u. for a potential in Hartree units.

We use a 1D Fourier transformation to obtain the passivant potential in reciprocal space:
\begin{eqnarray}
v_{\rm H} (G) &=& \frac{4\pi}{G} \int_0^\infty \sin{(Gr)} v_{\rm H} (r)r {\rm d}r \quad ,
\end{eqnarray}
with
\begin{equation}
v_{\rm H} (G=0) = 4\pi \int_0^\infty  v_{\rm H} (r)r^2 {\rm d}r \quad .
\end{equation}

\begin{figure}
\includegraphics[width=7.5cm]{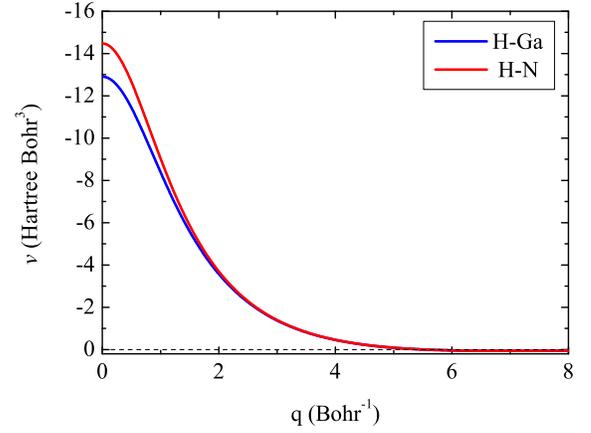}
\caption{Passivant screened pseudopotentials for Ga (H-Ga) and N (H-N) in reciprocal 
space.}
\label{hydro}
\end{figure}

In Figure~\ref{hydro} we show the pseudo-hydrogen passivant pseudopotentials in reciprocal 
space for the potential attached to Ga and to N. Both curves are simple and smooth, and 
deviate from each other only for small values of $q$, i.e. in their long range response, as expected. 

\begin{figure}
\includegraphics[width=7.5cm]{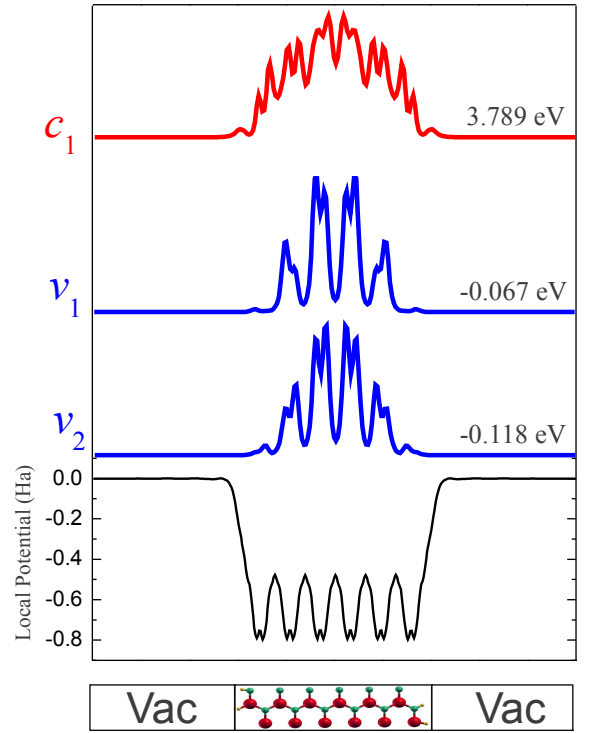}
\caption{Charge densities of one conduction band state, $c_1$ and 
two valence band states, $v_{1,2}$ of a passivated GaN slab calculated with 
the SEPM. Lower panel: atomic semiempirical pseudopotential across the slab. The 
quantities are averaged in successive planes perpendicular to 
the surface.}
\label{SEPM_layer}
\end{figure}

\textit{Step 3: Consistency checks on wave functions.} 
The accuracy of the passivation procedure has been checked by applying the SEPM to the same layer used for the LDA calculations. 
Figure~\ref{SEPM_layer} shows 
the potential and the charge densities of one conduction $c_1$ and two valence 
band $v_{1,2}$ states. The calculation are done with an energy cut-off of 30 Ry and a non-local 
part of the pseudopotential renormalized to the experimental bulk band gap. All spurious 
states are removed from the gap and the states are well localized inside the structure. The 
states $v_1$ and $v_2$ have the same profile than their counterparts LDA states. The small 
discrepancies can be attributed to the small dimensions of the layer, and gradually disappear for larger nanostructures.\cite{website}

\section{Electronic states of nanowires}
\label{nw}

To illustrate the performance of the SEPM in the study of nitride nanostructures, we present here
calculations of the electronic structure of
free-standing GaN NWs.
The NWs are assumed to be oriented along the $c$-axis and infinite
in length. We assume a hexagonal cross section, with $(1\bar{1}00)$
lateral faces. This is the most usual shape adopted by GaN NWs
grown by plasma-assisted molecular beam epitaxy technique.\cite{bert1,bruno1} This NW
faceting has also been supported by theoretical
calculations,\cite{nort1} showing a lower formation energy for
$(1\bar{1}00)$ than $(11\bar{2}0)$ surfaces.
The electronic states are obtained by solving
\begin{equation}\label{VNW}
\left[ -\frac{1}{2}\nabla^2+V_\mathrm{NW}(\bm{r})\right] \Psi_{k,n}(\bm{r}) =\varepsilon_n(k)\Psi_{k,n}(\bm{r}) \quad ,
\end{equation}
where the wire potential $V_\mathrm{NW}(\bm{r})$ is constructed as a superposition
of atomic pseudopotentials.

\begin{figure}
\includegraphics[width=7.0 cm]{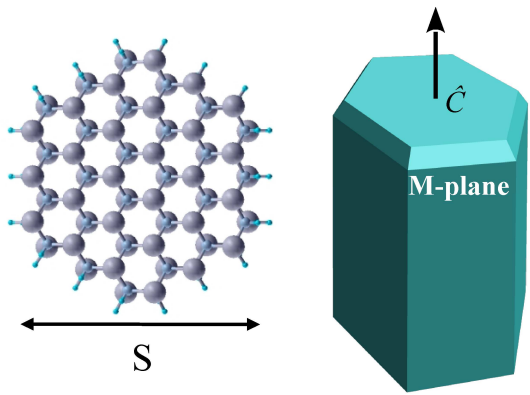}
\caption{
(Color online) Ball-and-stick model of a free-standing wurtzite
GaN nanowire, grown along the $[0001]$ direction. The lateral surfaces 
of the NW are $(1-100)$, and the corresponding surface atoms are passivated by
pseudo-hydrogen atoms.}
\label{SKETCH}
\end{figure}

In real free-standing NWs a surface reconstruction is expected that would minimizes
the total energy of the system. However,
in our calculations we will assume for simplicity that
the atomic arrangement corresponds everywhere to the perfect wurtzite crystal structure.
When building the potential $V_\mathrm{NW}(\bm{r})$
we have passivated the facets using the potential described in section \ref{surface}, which removes 
states from the gap in all the cases investigated.
Equation (\ref{VNW}) is solved by imposing artificial periodic boundary
conditions on the wire surrounded by $N_\mathrm{vac}$ layers of vacuum. This transforms
Eq. (\ref{VNW}) into a Bloch-periodic band-structure problem with
a large simulation cell, solved here by expanding $\Psi_{k,n}(\bm{r})$ in plane
waves. We use a sufficiently large wire-wire separation $N_\mathrm{vac}$,
so that the solutions become independent of $N_\mathrm{vac}$.
A sketch of the modeled NW
is shown in Fig. \ref{SKETCH}.
In the following, the term \textit{size} refers to the largest lateral dimension, as shown in 
in Fig. \ref{SKETCH} and is labeled as $S$.
We have considered  a range of sizes $S$ from 1.5
to 6.5 nm, which corresponds to a number of atoms varying
between 100 to 1500.
The matrix diagonalization corresponding to the problem Eq. (\ref{VNW})
can be solved using the folded spectrum method.\cite{fsm}
The computational times, within the SEPM, are significantly shorter than
they would be in a corresponding self-consistent \textit{ab initio} calculation.

Here, we will focus on the analysis of the zone center ($k=0$) NW electronic states.
These states are naturally classified into valence band (VB, $v$) and conduction band (CB, $c$) NW states, and in
each case they will be indexed in ascending order, as $(v_1, v_2, \dots)$ and
$(c_1, c_2, \dots)$, according to their separation in energy
from the respective $\bm{k}=0$ bulk band edges $E_v$ and $E_c$.
Any of the considered states exhibits Kramer's double degeneracy.
When analyzing the corresponding  charge densities (here meaning,
i.e., the (vertically-averaged) squared wave functions),
an envelope pattern is recognized in the cross-section, which corresponds
very closely to the radial pattern of the Bessel functions $J_n(x)$. We
find convenient to incorporate this symmetry information to the labeling of the states
by adding the notation $s$ (meaning: \textit{approximately the same symmetry as $J_0(x)$}),
$p$ (\textit{approximately the same symmetry as $J_1(x)$}), etc...

In order to further characterize the symmetry of the NW states
$\Psi_{k=0,n}$, we can calculate their projections on the bulk band states,
as explained in Ref.~\onlinecite{baskoutas10}. In particular,
in the analysis presented below, we have considered the projections
$$
P_{n}(\gamma)=\langle \Psi_{ \gamma } | \Psi_{k=0,n} \rangle
\quad
$$
where $\Psi_{ \gamma }$ are the near-gap
 $\Gamma$ ($\bm{k} = \bm{0}$)
bulk states labeled schematically by their
symmetries: $\gamma=\Gamma_{7,c}$ (bottom of the CB)
and $\gamma=A(\Gamma_9)$, $B(\Gamma_{7,+})$, $C(\Gamma_{7,-})$.\cite{vurgaftman2003} 

\begin{figure}
\includegraphics[width=8.cm]{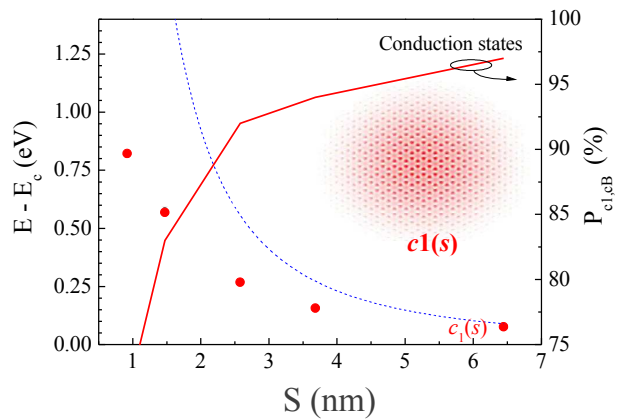}
\caption{
(Color online) On the left scale it is indicated the confinement energy of the CB NW states 
as a function of the NW size. The blue dashed line is the fit obtained with Eq. (\ref{fiteq}) in the 
text, whereas the red dots represents
the confinement energy of the lowest CB NW.
The solid line refers to the right scale and it represents the projection 
of the $c_1$ wave function onto the
CB edge bulk state.
The charge density of the CB state $c_1(s)$
for $S=6.5$ nm is shown as an inset.}
\label{cond}
\end{figure}	

Figure~\ref{cond} shows the confinement energies (i.e., the energies
referred to the bulk CB edge energy)
of the CB NW states as a function of the NW size.
It is also shows the projection of the lowest energy state $c_1$
onto the CB bulk state $\Psi_{ \Gamma_{7,c}}$, 
revealing that $c_1$ has between 65 \% and 97 \% conduction band character.
Additionally, the charge density of $c_1$ exhibits a clear $s$-type envelope for
all the sizes explored here, in agreement with expectations from the EMA. 
On the other hand, we have fitted the
energy of the $c_1(s)$ state to the function
\begin{equation}
\varepsilon(S)-E_c =  a\frac{1}{S^b},
\label{fiteq}
\end{equation}
and have found the values $a=1.28\pm0.01$ and $b=1.62\pm0.02$. 
The fitted value of the parameter $b$ differs
significantly from the prediction of the single-band EMA  ($b=2$), which is
represented by a dotted line in Fig.~\ref{cond}.
We conclude that, even for the case of the almost isolated conduction band,
the confinement imposed by the nanowire geometry,
while leading to an overall symmetry of the CB states similar to that predicted by the EMA,
leads to a remarkably different energy dispersion as a function of NW size.

\begin{figure}
\includegraphics[width=7.5cm]{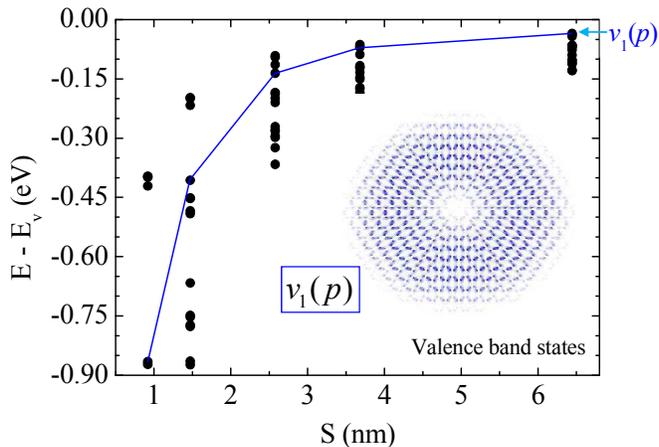}
\caption{
(Color online) Size dependence of the confinement energy of the first VB NW states as a 
function of the NW size.
The charge density of the
first VB state for a NW size of 6.5 nm, $v_1(p)$, is shown as an inset.
The solid line connects the size dispersion of the hybrid $A$-$B$ state with $p$-type
envelope symmetry, which is discussed in the main text. 
}
\label{vale}
\end{figure}

\begin{figure*}
\includegraphics[width=15 cm]{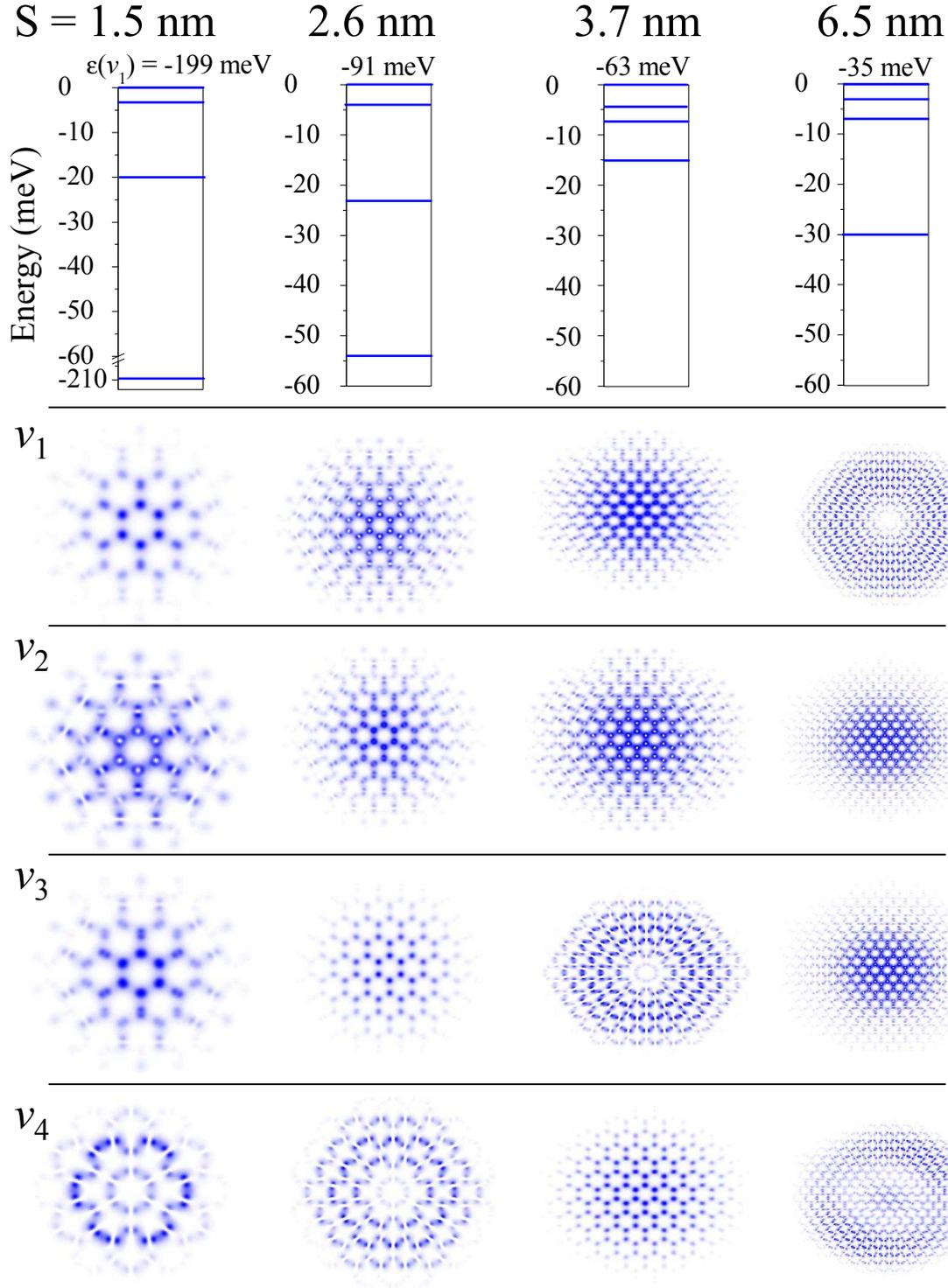}
\caption{
(Color online)
The upper panel gives the single particle energies for the VB NW  states, 
$v_1$, $v_2$, $v_3$, and $v_4$, relative to the highest VB state $\varepsilon(v_1)$ for various NW sizes. The 
corresponding charge densities are displayed in the lower panels.}

\label{evo}
\end{figure*}

We now turn to the analysis of the VB electronic structure of the NW.
Figure~\ref{vale} shows an overall view of the confinement effects
in the VB by displaying the highest energy levels 
for every NW size.
In order to give further insight into the complex evolution of the VB states with sizes, 
we have represented in Fig.~\ref{evo} the charge densities of the first four VB states 
for various NW sizes. 
A first examination reveals that the wave functions are well confined inside
the NWs, and only for some states in the smallest NW show a slight state localization 
on surface atoms.
In particular, the analysis of the charge density of the highest VB state $v_1$
shows that its envelope symmetry changes suddenly from $p$-type for the largest NW to $s$-type for
smaller NWs. For the size 2.6 nm and below, the first three VB states
show a $s$-type envelope, whereas the VB state with $p$-type envelope is moved to the fourth
place in the energy spectrum. 

\begin{table}
\begin{tabular}{lcccc}
\hline
\hline
S(nm)   &   1.5        &    2.6       &   3.7       &   6.5        \\
\hline
$v_1$   &  (3, 40, 46) &  (76, 8, 6)  & (79, 9, 5)  & (50, 47, 0)  \\
$v_2$   &  (60, 7, 5)  &  (7, 68, 17) & (9, 75, 11) & (82, 12, 3)  \\
$v_3$   &  (4, 29, 53) &  (3, 8, 85)  & (48, 45, 1) & (18, 76, 3)  \\
$v_4$   &  (41, 42, 0) &  (45, 45, 0) & (3, 6, 89)  & (38, 42, 16) \\
\hline
\hline
\end{tabular}
\caption{Projections
($P_{v_i}(A)$, $P_{v_i}(B)$, $P_{v_i}(C)$) of the upper VB NW states 
$v_i$ ($i=1,2,3,4$), onto the bulk VB states at $\Gamma$, ($A$, $B$, $C$).
 }
\label{tabla}
\end{table}

The above study must be complemented with the analysis of the projections of
the NW VB states onto the bulk states $A$,
$B $ and $C $, at the $\Gamma$ point, which
will give us additional information about the symmetry of the states.
In Table.~\ref{tabla} we have summarized the values of the
corresponding projections ($P_{v_i}(A)$, $P_{v_i}(B)$ and
$P_{v_i}(C)$), for the same set of states as in
Fig.~\ref{evo}. For the largest nanowire shown here,
the composition of
the topmost state, which has envelope $p$ symmetry,
is almost equally divided between the bulk states
$A$ and $B$.
The next two VB states, $v_2$ and $v_3$, of $s$-symmetry, clearly
show a dominant contribution from bulk bands $A$  ($P_{v_2}(A)=82 \%$) and $B$
($P_{v_3}(B)=76 \%$), respectively. Neither of these
states has a significant contribution from the $C$-band. When the
nanowire size is reduced to 3.7 nm,  we observe that the two highest VB states
have now dominant contributions from $A$- and $B$-bands, with $s$ symmetry, 
whereas the  mixed $A$-$B$  state becomes now the third state, $v_3(p)$.  
 If one further reduces the NW size to
2.6 nm and below, the mixed $A$-$B$ state now becomes $v_4(p)$, whereas
the first three states show a high degree of $A$-$B$-$C$ mixing.
Thus, our detailed analysis of the NW wave functions as a
function of size has allowed us to identify a state with $p$-type
envelope symmetry which is essentially an equal-weight
 mixture of bulk bands $A$ and $B$. In Fig.~\ref{vale} we have traced a line connecting these states.
The change in the symmetry of the highest VB state as a function of size
is just one of the consequences of the  nontrivial interplay between
symmetry mixing, spin-orbit coupling and confinement effects on the NW 
valence band electronic structure. This is in contrast to the simpler situation of the 
CB electronic structure.
This unconventional trend of the nanowire electronic
structure has also been reported by calculations using the
tight-binding method.\cite{Persson2008,molina2012}
Moreover, studies in other
wurtzite systems, as the one made in ZnO nanocrystals by S.
Baskoutas and G. Bester \cite{baskoutas10} shows also that the
highest state of the valence band is characterized by a $p$-type
envelope.

\section{Conclusion}

We have followed the idea \cite{wang95,fu97b} to use the screened effective potential from a 
DFT bulk calculation to extract a spherically averaged atomic pseudopotential. We have derived 
the analytic connection between the screened DFT effective potential and the atomic quantities in 
the case of wurtzite bulk unit cells. This allowed us to derive atomic semi-empirical pseudopotentials 
for nitride materials. The pseudopotentials are non-local, in the same way as DFT norm-conserving 
pseudopotentials, and reproduce the DFT results for the eigenvalues of bulk up to a few meV. Our 
potentials are not aimed at the calculation of total energies but at the calculations of the 
eigenvalues and eigenfunctions around the band gap, which gives some leverage on the reduction of the energy cut-off, compared to DFT calculations.  
The advantage of the method compared to the original empirical pseudopotential 
approach (see Ref.~\onlinecite{Cohen70} and references therein) resides in the direct link 
to ab-intio calculations and the ensuing rather simple fitting procedure of a curve through a dense set of data points. 
The transferability of the potentials are good for the situations studied here but are expected to 
degrade when charge transfer significantly deviates from the bulk situation. This represents a general 
limitation of the method that we, however, palliate for the case of a free surface by deriving effective passivant potentials (see next paragraph).
The defective band gap inherited from DFT is correct by a slight adjustment of the non-local channel 
of the potential which has only a very marginal effect on the wave functions, as shown previously \cite{wang95,fu97b,li05a} .
The derived method should be applicable to other materials. Eventually, the method can be built on the basis
of GW calculations. However, full self-consistent GW calculations are still prohibitive in the applications
to extended systems and only work for small molecules.\cite{rinke12}

In the second part of this work, we have established a simple and direct 
methodology to obtain the passivating pseudo-hydrogen pseudopotentials, essential for
the treatment of free surfaces in nanowires or nanoparticles. We 
thereby use the difference between the screened effective DFT potentials of a passivated structure and
the semiempirical pseudopotential of the corresponding unpassivated 
structure. The derivation is done in real space and we find 
that a cubic spline interpolation up to a cut-off 
radius $R_C$, connected to a Yukawa potential leads to satisfactory results.

The suitability of the semi-empirical pseudopotential method and of our passivation procedure has been explored in
GaN nanowires. Due to the high confinement of NW sizes below 10 nm, the use of accurate atomic
potentials is unavoidable, as continuous methods are expected to fail in the description of such systems. We have found
that, while the conduction band states follow a behavior compatible with the EMA predictions, the valence band states
exhibit a complex evolution as a function of the NW size. We find drastic variations in the character of the
wave functions with the NW size, which will influence optical properties such as the polarization of the photoluminescence.

\section{Acknowledgements}

A. Molina-S\'{a}nchez thanks the Deutscher Akademischer Austausch Dienst (DAAD) for 
funding the research stay at the Max-Planck Institute f\"{u}r 
Festk\"{o}rperforschung in Stuttgart (Germany), and the group of Gabriel Bester for its
kind hospitality.


\begin{thebibliography}{1}

\bibitem{hohenberg64a} P. Hohenber and W. Kohn, Phys. Rev. \textbf{136}, B864 (1964).

\bibitem{kohn65a} W. Kohn and L. J. Sham, Phys. Rev. \textbf{140}, A1133 (1965).

\bibitem{wang95} L.-W. Wang and A. Zunger, Phys. Rev. B \textbf{51}, 17398 (1995).

\bibitem{bester09} G. Bester, J. Phys. Condens. Mat. \textbf{21}, 023202 (2009)

\bibitem{fu97b} H. Fu and A. Zunger, Phys. Rev. B \textbf{55}, 1642 (1997).

\bibitem{bellaiche96} L. Bellaiche, S.-H. Wei, and Z. Zunger, Phys. Rev. B, \textbf{54}, 17568 (1996).

\bibitem{grundmann} D. Fritsch, H. Schmidt, and M. Grundmann, Phys. Rev. B \textbf{67}, 235205 (2003).

\bibitem{williamson00} A. J. Williamson, L.-W. Wang, and A. Zunger, Phys. Rev. B \textbf{62}, 12963 (2000).

\bibitem{graf07} P. A. Graf, K. Kim, W. B. Jones, and L.-W. Wang, J. Comp. Phys. \textbf{224}, 824 (2007).

\bibitem{Goldberger} J. Goldberger, R. He, Y. Zhang, S. Lee, H. Yan, H.-J. Choi, and P. Yang.
Nature \textbf{422}, 599 (2003).

\bibitem{wu} J. Wu. J. Appl. Phys. \textbf{106}, 011101 (2009).

\bibitem{gonze09} All DFT calculations are done with ABINIT. X. Gonze, B. Amadon, P.-M. Anglade, J.-M. Beuken, 
F. Bottin, P. Boulanger, F. Bruneval, D. Caliste, R. Caracas, M. Cote, 
T. Deutsch, L. Genovese, Ph. Ghosez, M. Giantomassi, S. Goedecker, 
D.R. Hamann, P. Hermet, F. Jollet, G. Jomard, S. Leroux, M. Mancini, 
S. Mazevet, M.J.T. Oliveira, G. Onida, Y. Pouillon, T. Rangel, 
G.-M. Rignanese, D. Sangalli, R. Shaltaf, M. Torrent, M.J. Verstraete, 
G. Zerah, J.W. Zwanziger, Computer Phys. Commun. \textbf{180}, 2582 (2009).

\bibitem{Parr} R. G. Parr and W. Yang, \textit{Density-Functional 
Theory of Atoms and Molecules}, (Oxford University Press, 1989).

\bibitem{Grabowski2001} S. P. Grabowski, M. Schneider, H. Nienhaus, W. Monch, R. Dimitrov,
O. Ambacher, and M. Stutzmann. Appl. Phys. Lett. \textbf{78}, 2503 (2001). 

\bibitem{zhang1993} S. Zhang, C. Yeh, A. Zunger, Phys. Rev. B, \textbf{48}, 11204 (1993).

\bibitem{mader} K. Mader and A. Zunger, Phys. Rev. B, \textbf{50}, 17393 (1994).

\bibitem{bandgap} L. J. Sham and M. Schl\"{u}ter, Phys. Rev. Lett. \textbf{51}, 1888 (1983).

\bibitem{monemar} B. Monemar, J. Phys: Condens. Matt. \textbf{13} 7011 (2001).

\bibitem{wang96a} L.-W. Wang and A. Zunger, Phys. Rev. B \textbf{53}, 9579 (1996).

\bibitem{franceschetti99} A. Franceschetti, H. Fu, L.-W. Wang, and A. Zunger, Phys. Rev. B \textbf{60}, 1819 (1999).

\bibitem{puzder03} A. Puzder, A. J. Williamson, F. A. Reboredo, and G. Galli, Phys. Rev. Lett. \textbf{91}, 157405 (2003).

\bibitem{puzder04a} A. Puzder, A. J. Williamson, F. Gygi, and G. Galli, Phys. Rev. Lett. \textbf{92}, 217401 (2004).

\bibitem{puzder04b} A. Puzder, A. J. Williamson, N. Zaitseva, G. Galli, L. Manna, and A. P. Alivisatos, Nano Lett. \textbf{4}, 2361 (2004).

\bibitem{califano03} M. Califano, G. Bester, and A. Zunger, Nano Lett. \textbf{3}, 1197 (2003).

\bibitem{baskoutas11} S. Baskoutas and G. Bester, J. Phys. Chem. C \textbf{115}, 15862 (2011).

\bibitem{xu93} Y. N. Xu and W. Y. Ching, Phys. Rev. B \textbf{48}, 4335 (1993).

\bibitem{huang05} X. Huang, E. Lindgren, and J. R. Chelikovsky, Phys. Rev. B \textbf{71}, 165328 (2005).

\bibitem{reboredo01} F. A. Reboredo and A. Zunger, Phys. Rev. B \textbf{63}, 235314 (2001).

\bibitem{distances} The distances are $d_{\rm{H-N}}=0.541$ and $d_{\rm{H-Ga}}=0.786$, in units of 
GaN bulk distance, $d_{\rm{Ga-N}}=uc$.

\bibitem{molina09} A. Molina, A. Garc\'{i}a-Crist\'{o}bal, and A. Cantarero, Microelectr. J. \textbf{40}, 418 (2009).

\bibitem{website} All the pseudopotentials can be found, tabulated, at the following 
website: http://www.fkf.mpg.de/bester/ and in the suplementary material.

\bibitem{bert1} K. A. Bertness, A. Roshko, L. M. Mansfield, T. E. Harvey, and N. A. Sanford, J. Cryst. Growth \textbf{300}, 94 (2007).

\bibitem{bruno1} R. Songmuang, O. Landr\'{e}, and B. Daudin, Appl. Phys. Lett. \textbf{91}, 251902 (2007).

\bibitem{nort1} J. E. Northrup and J. Neugebauer, Phys. Rev. B \textbf{53}, R10477 (1996).

\bibitem{fsm} L.-W. Wang and Z. A., J. Chem. Phys. \textbf{100}, 2394 (1994).

\bibitem{baskoutas10} S. Baskoutas and G. Bester, J. Phys. Chem. C \textbf{114}, 9301 (2010).

\bibitem{vurgaftman2003} I. Vurgaftman and J. R. Meyer, J. Appl. Phys. \textbf{94}, 3675 (2003).

\bibitem{Persson2008} M. P. Persson and A. Di Carlo, J. Appl. Phys. \textbf{104}, 073718 (2008).

\bibitem{molina2012} A. Molina-S\'{a}nchez and A. Garc\'{i}a-Crist\'{o}bal, J. Phys.: Condens. Matter \textbf{24}, 295301 (2012).

\bibitem{Cohen70} M. L. Cohen and V. Heine. \textit{The fitting of pseudopotentials
to experimental data and their subsequent application}. Vol. 24 of Solid State Physics pp. 37-248 (Academic Press, 1970).

\bibitem{li05a} J. Li and L.-W. Wang, Phys. Rev. B \textbf{72}, 125325 (2005).

\bibitem{rinke12} F. Caruso, P. Rinke, X. Ren, M. Scheffler, and A. Rubio, Phys. Rev. B 86, 081102(R) (2012).

\end{thebibliography}
\end{document}